\documentclass[aps,11pt]{revtex4}
\usepackage{graphicx}
\usepackage{amssymb}
\begin{document}
\title{A two-step MaxLik-MaxEnt strategy to infer photon distribution
from on/off measurement at low quantum efficiency}
\author{Andrea R. Rossi and Matteo G. A. Paris}
\affiliation{Dipartimento di Fisica dell'Universit\`a degli Studi di
Milano, Italia}
\date{\today}
\begin{abstract}A method based on Maximum-Entropy (ME) principle to
infer photon distribution  from \emph{on/off} measurements
performed with \emph{few} and \emph{low} values of quantum
efficiency is addressed. The method consists of two steps: at
first some moments of the photon distribution are retrieved from
\emph{on/off} statistics using Maximum-Likelihood estimation, then
ME principle is applied to infer the quantum state and, in turn,
the photon distribution. Results from simulated experiments on
coherent and number states are presented. \end{abstract}
\maketitle
\section{Introduction}
Besides fundamental interest, the reconstruction of the photon
distribution of an optical signal $\rho$, plays a major role in
high-rate quantum communication schemes based on light beams
\cite{rmp}, and is required for implementations of linear-optics
quantum computing \cite{lcomp}.  Effective photon counters have
been indeed developed, though their  current operating conditions
are still extreme \cite{xxx}.  At present, the most convenient
method to infer photon distribution is perhaps quantum tomography
\cite{revt}, which have been applied to several quantum states
\cite{mun,raymerLNP} with reliable statistics. However, the
tomography of a state needs the implementation of homodyne
detection, which in turn requires the appropriate mode matching of
the signal with a suitable local oscillator at a beam splitter. As
a matter of fact, quantum tomography has been developed to gain a
complete characterization of the signal under investigation, and
may not be suited in case where we are interested only in
obtaining partial information, as for example the photon
distribution \cite{added}. \par An alternative approach, based on
an array of avalanche \emph{on/off} photodetectors with different
quantum efficiencies has been suggested \cite{mogy} and
demonstrated with fiber-loop detectors \cite{olom}. In this
scheme, repeated preparations of the signal are revealed at
different quantum efficiencies, and the resulting \emph{on/off}
statistics is then used to reconstruct the photon distribution
through Maximum-Likelihood estimation. The statistical reliability
of the method has been also analyzed in details \cite{qph}.
\par
In this paper we want to further reduce the experimental
requirements. We assume that avalanche photodetection may be
performed only with low values of the quantum efficiency and, in
addition, that only few of those values are available. Then we
analyze whether the photon distribution may be inferred from this
scarce piece of information. We found that the use of Maximum
Entropy principle, together with Maximum Likelihood estimation of
moments of the distribution, provides an effective method to
obtain the full photon distribution. \par In Section \ref{s:due}
we describe in details the two-step MaxLik-MaxEnt method, whereas
in Section \ref{s:tre} results from numerically simulated
experiments on coherent and number states are reported. Section
\ref{s:out} closes the paper with some concluding remarks.
\section{Photon distribution using low efficiency avalanche detectors}\label{s:due}
Maximum Entropy (ME) principle \cite{jaynes} is a powerful tool to
infer the density matrix of a quantum state when only a
\emph{partial} knowledge about the system has been obtained from
measurements. Let us call  $\mathcal{O}$ the set of independent
operators $\mathrm{O}_\nu $ summarizing the measurements performed
on a given systems. $\mathcal{O}$ is called the \emph{observation
level} \cite{buzek} gained from the measurements.  The ME
principle states that the density matrix compatible with a given
$\mathcal{O}$ is that maximizing the Von Neumann entropy while
reproducing the known values of the operators in $\mathcal{O}$. In
order to enforce this condition the inferred density operator will
depend on a number of parameters whose values must be properly
determined. ME principle is a way of filling our ignorance about
the system, without assuming more than what we have learned from
measurements. The actual form of the inferred state heavily
depends on the observation level $\mathcal{O}$: it can range from
a (nearly) complete knowledge of the state, as it happens for
quantum tomography, to the (nearly) complete ignorance. An example
of the latter case is the knowledge of the mean photon number
alone, for which ME principle yields a thermal state. In general,
the ME density operator $\varrho_{ME}$ that estimates the signal $\rho$ 
for a given observation level is given by
\begin{eqnarray} \label{rhomaxent}
\varrho_{ME}
=Z^{-1}\exp\left\{-\sum_{\nu} \lambda_{\nu}\mathrm{O}_{\nu}
\right\}\:,
\end{eqnarray}
where $Z=\hbox{Tr}\left[\exp\left\{-\sum_{\nu} 
\lambda_{\nu}\mathrm{O}_{\nu}
\right\}\right]$ is the partition
function, and the coefficients $\lambda_{\nu}$ are Lagrange
multipliers, to be determined by the constraints 
$$ \hbox{Tr}\left[\varrho_{ME} \:\mathrm{O}_{\nu} \right]\equiv
-\partial_{\lambda_{\nu}}\log{Z} = \langle \mathrm{O}_{\nu}
\rangle\,,$$ where $\langle \mathrm{O}_{\nu} \rangle$ are the
expectation values obtained from the measurements. Suppose now
that we would like to measure the photon statistics of a given
signal $\rho$ using only avalanche photodetectors with
efficiencies $\eta_\nu$, $\nu=1,...,N$, \emph{i.e.} we perform
\emph{on/off} measurements on the signal with $N$ different values
of the quantum efficiency. The statistics of the measurements is
described by the probability of the \emph{off} events when the
quantum efficiency is $\eta_\nu$
\begin{eqnarray}\label{p_nu_def}
p_\nu &=&\hbox{Tr}\left[\rho\:\Pi_\nu \right] \nonumber \\
&=&\sum_{n=0}^{\infty}(1-\eta_{\nu})^n \rho_{n}\:.
\end{eqnarray}
In Eq. (\ref{p_nu_def}) $\rho_n \equiv \langle n \vert \rho \vert
n \rangle$ and $\Pi_\nu\equiv \Pi_{\rm{off}}(\eta_\nu)$, where the
probability measure (POVM) of the measurements is given by
\begin{eqnarray} \label{povm}
\Pi_{\rm{off}}(\eta_\nu)
&=&\sum_{n=0}^{\infty}(1-\eta_{\nu})^n \: \vert n \rangle
\langle n \vert \nonumber \\
\Pi_{\rm{on}}(\eta_\nu)&=&\mathbb{I}-\Pi_{off}(\eta_\nu)\:.
\end{eqnarray}
\par
From the ME principle we know that the best state we can infer is
given by (\ref{rhomaxent}) that, in this case, reads as follows
\begin{eqnarray} \label{maxent_rho}
\varrho_{ME}=
Z^{-1}\exp\left\{-\sum_{\nu=1}^N \lambda_\nu
\Pi_\nu\right\}.
\end{eqnarray}
Explicit equations for the $\lambda_\nu$ are obtained expanding
the above formulas in the Fock basis; we have
\begin{eqnarray}\label{x1}
p_{\nu}= \frac{
\sum_{n=0}^{\infty} (1-\eta_\nu)^n \: \exp \left\{ -\sum_{\mu=1}^{N}
\lambda_{\mu} \left(1-\eta_{\mu}\right)^n \right\}
}{\sum_{n=0}^{\infty} \exp \left\{ -\sum_{\mu=1}^{N}
\lambda_{\mu} \left(1-\eta_{\mu}\right)^n \right\}}
\:.
\end{eqnarray}
Eq. (\ref{x1}) can be solved numerically in order to determine the
coefficients $\lambda_\nu$ and in turn the ME density operator. In
the following we consider situations where the experimental
capabilities are limited. We suppose that \emph{on/off}
measurements can be taken only at \emph{few} and \emph{low} values
of the quantum efficiencies. In this case the statistics
(\ref{p_nu_def}) can be expanded as
\begin{eqnarray}
p_\nu &=&
\sum_{n}\left\{\rho_{n}-n\eta_{\nu}\rho_{n}+\frac{1}{2}n(n-1)\eta_{\nu}^2
\rho_{n}+\right. \nonumber \\
&\hbox{}&\left.-\frac{1}{6}n(n-1)(n-2)\eta_{\nu}^3
\rho_{n}+\ldots\right\}\:.
\end{eqnarray}
Summing the series we have
\begin{eqnarray} \label{p_nu_approx}
p_\nu
&=&1-\eta_{\nu}N_1+\frac{1}{2}(N_2-N_1)\eta_{\nu}^2+ \nonumber \\
&\hbox{}& -\frac{1}{6}(N_3-3N_2+2N_1)\eta_{\nu}^3+\ldots \nonumber \\
&=&1- N_1 (\eta_{\nu} + \frac12 \eta_{\nu}^2+ \frac13 \eta_{\nu}^3+ \ldots)
+ N_2 (\frac12 \eta_{\nu}^2 + \frac12 \eta_{\nu}^3 + \ldots )
\nonumber \\
&\hbox{}&- N_3 (\frac16 \eta_{\nu}^3+
\ldots) + \ldots
\end{eqnarray}
where
\begin{eqnarray}
N_k=\overline{n^k}=\hbox{Tr}\left[\rho\: (a^{\dag} a)^k\right],
\end{eqnarray}
are moments of the photon distribution. By inversion of Eq.
(\ref{p_nu_approx}), upon a suitable truncation, we retrieve the
first moments of the distribution from the \emph{on/off}
statistics at low quantum efficiency.
\par
The most effective technique to achieve the inversion of the above
formula is Maximum-Likelihood. The Likelihood of the \emph{on/off}
measurement is given by
\begin{eqnarray} \label{likelyhood}
L=\prod_{\nu=1}^N
p_{\nu}^{n_{\nu}}\left(1-p_{\nu}\right)^{\mathcal{N}_{\nu}-n_{\nu}},
\end{eqnarray}
where $n_{\nu}$ and $\mathcal{N}_{\nu}$ are respectively the
number of \emph{off} events, and the total number of measurement
when using a detector with efficiency $\eta_{\nu}$. In practical
calculations it is more convenient to use the logarithm of
(\ref{likelyhood}), that reads as follows:
\begin{eqnarray} \label{loglik}
\mathcal{L} &\equiv& \log{L} \nonumber \\
&=&
\sum_{\nu=1}^N\Big(n_{\nu}\log{p_{\nu}}+
(\mathcal{N}_{\nu}-n_{\nu})\log{(1-p_{\nu})}\Big).
\end{eqnarray}
Without loss of generality we can set
$\mathcal{N}_{\nu}=\mathcal{N}$, $\forall \nu$ and divide
(\ref{loglik}) by $\mathcal{N}$, obtaining:
\begin{eqnarray} \label{logliknorm}
\frac{\mathcal{L}}{\mathcal{N}}=\sum_{\nu=1}^{N}\Big(f_{\nu}\log{p_{\nu}}+
(1-f_{\nu})\log{(1-p_{\nu})}\Big),
\end{eqnarray}
where $f_{\nu}$ are the experimental frequencies of the \emph{off}
events. Substituting (\ref{p_nu_approx}) in (\ref{logliknorm}) we
find an expression for the renormalized likelihood as a function
of the moments $N_k$.
\par
Maximization of (\ref{logliknorm}) over the parameters $N_k$
leads to the following set of optimization equations:
\begin{eqnarray} \label{grad}
\frac{1}{\mathcal{N}}\frac{\partial \mathcal{L}}{\partial N_k}
\equiv
\sum_{\nu=1}^N\left(\frac{f_{\nu}}{p_{\nu}}-
\frac{1-f_{\nu}}{1-p_{\nu}}\right)\frac{\partial
p_{\nu}}{\partial N_k}=0.
\end{eqnarray}
If we stop the expansion in (\ref{p_nu_approx}) at the second
order (see below) the derivatives in (\ref{grad}) take the form:
\begin{eqnarray}
\frac{\partial p_\nu}{\partial
N_1}&=&-(\eta_\nu+\frac{\eta_\nu^2}{2}) \nonumber \\
\frac{\partial p_\nu}{\partial N_2}&=&\frac{\eta_\nu^2}{2}.
\end{eqnarray}
The system (\ref{grad}) can be easily solved numerically given the
\emph{on/off} statistics as well as the number and the values of
the quantum efficiencies used during the experiments.
\par
After having determined the first moments, the density matrix of
the signal, according to Maximum Entropy principle, is given by
\begin{eqnarray}
\varrho_{ME}=Z^{-1}
\exp\left\{-\sum_{\nu}^{2} \lambda_\nu (a^\dag
a)^\nu\right\}
\end{eqnarray}
with $\lambda_\nu$ to be determined as to satisfy
\begin{eqnarray} \label{conditions}
N_k&=&\hbox{Tr}\left[ \varrho_{ME}\:(a^\dag a)^k \right]
\nonumber \\
&=& \frac{
\sum_{n=0}^{\infty} n^k \: \exp \left\{ -\sum_{\nu=1}^{2}
\lambda_{\nu} n^\nu \right\}
}{\sum_{n=0}^{\infty} \exp \left\{ -\sum_{\nu=1}^{2}
\lambda_{\nu} n^\nu \right\}}\:.
\end{eqnarray}
Notice that the unknowns $\lambda_\nu$ are contained also
in the denominator and that a suitable truncation should be
adopted (which can be easily determined by imposing normalization
on the ME density matrix).
\section{Simulated experiments} \label{s:tre}
\begin{figure}[h!]
\begin{center}
\includegraphics[width=0.35\textwidth]{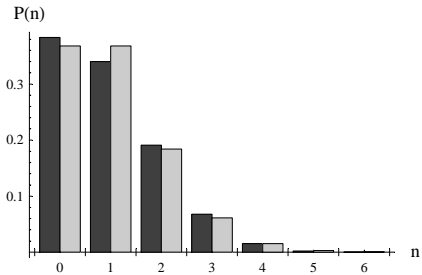}
\end{center}
\caption{Reconstruction of the photon distribution of a coherent
state with $|\alpha|^2=1$ (black), and comparison with the
theoretical values (light gray). Simulated \emph{on/off}
measurements have been performed with $N=5$ values of $\eta$
ranging between 1\% and 5\%. The number of experimental measures
is $10^6$ for each $\eta$. Fidelity of the reconstruction is
larger than $F=99\%$.} \label{f:fig1}
\end{figure}

\begin{figure}[h]
\begin{center}
\includegraphics[width=0.35\textwidth]{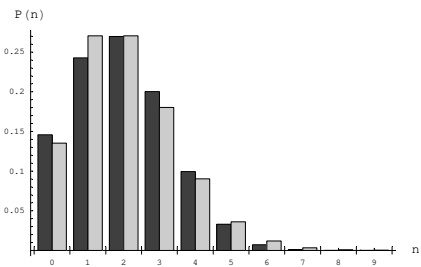}
\end{center}
\caption{Reconstruction of the photon distribution of a coherent
state with $|\alpha|^2=2$ (black) and comparison with the
theoretical values (light gray). Simulations parameters are the
same as in Fig. \ref{f:fig1}. Fidelity of the reconstruction is
larger than $F=99\%$.} \label{f:fig2}
\end{figure}

\begin{figure}[h]
\begin{center}
\includegraphics[width=0.35\textwidth]{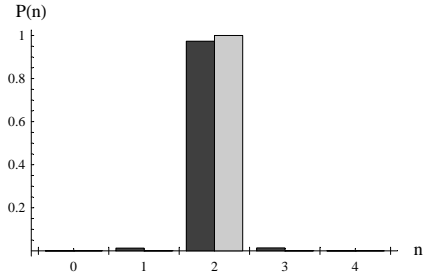}
\end{center}
\caption{Reconstruction of the photon distribution of the number
state $\vert 2 \rangle$ (black) and comparison with the
theoretical values (light gray). Simulations parameters are the
same as in Fig. \ref{f:fig1}. Fidelity of the reconstruction is
$F=98.5 \%$.} \label{f:fig3}
\end{figure}

We have performed simulated experiments of the whole procedure on
coherent and number states. For this kind of signals the first two
moments are sufficient to obtain a good reconstruction via ME
principle because their Mandel parameter $Q=-1+(N_2-N_1^2)/N_1$
is less than or equal
to the average photon number $N_1$ \cite{buzek}, while squeezed
states, for which $Q>N_1$, are ruled out, requiring the knowledge
of a considerably larger set of moments (in principle, all of them).
A number $N=5$ of values of the quantum efficiency ranging from $1\%$ to
$5\%$ have been enough in order to achieve good reconstructions.
Our results are summarized in Figs. \ref{f:fig1}-\ref{f:fig3}.
Notice that a faithful photon statistics retrieval needs a
sufficiently accurate knowledge of the $p_{\nu}$'s: our results
are obtained using ${\cal N}=10^6$ observations for each $\eta_\nu$.
This condition can be relaxed if we increase the number $N$ of
probabilities measured, but their range of values should be kept
narrow because Eq. (\ref{p_nu_approx}) must hold.
\par
\par
\begin{figure}[h]
\begin{center}
\includegraphics[width=0.35\textwidth]{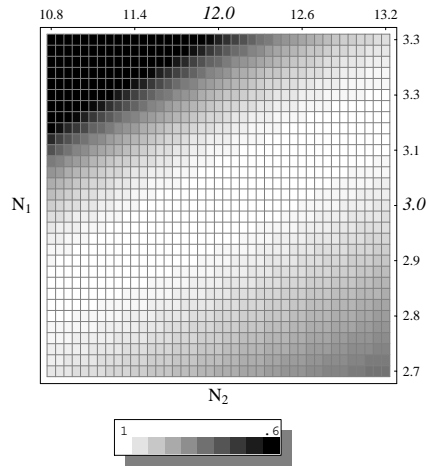}
\end{center}
\caption{Fidelity of the photon distribution reconstruction for a
coherent state $|\alpha|^2=3$, as a function of the average photon
number $N_1$ and its second moment $N_2$. The true values for the
two parameters are printed in italics. Simulations parameters are
the same as in Fig. \ref{f:fig1}. Notice that the first two
entries in the top line are set to 0 because, in these cases, the
Mandel parameter $q$ is smaller than -1, so they are not physical
states.} \label{f:fig4}
\end{figure}

\noindent
Finally in Fig. \ref{f:fig4} we check the robustness of the
ME inference, against errors in the knowledge of the parameters $N_1$
and $N_2$ that may come from ML estimation in the first step.
The quality of the reconstruction has been assessed
through fidelity
$$ F=\sum_n \sqrt{q_n\, p_n}\:,$$
between the inferred
$q_n = \langle n| \varrho_{ME}|n\rangle$
and the true
$p_n = \langle n| \rho|n\rangle$
photon  distribution. As it is apparent from the plot the reconstruction's
fidelity remains large if the relative errors on both parameters are about
$\pm 5\%$.
\section{Conclusions} \label{s:out}
In this paper we have shown that the photon distribution of a
light signal $\rho$, with Mandel parameter lower than or equal to
its average photon number, can be reconstructed using few
measurements collected by a low efficiency avalanche
photodetector. The \emph{on/off} statistics is used in a two steps
algorithm, consisting in retrieving the first two moments of the
photon distribution via a Maximum-Likelihood estimation, and than
inferring the diagonal entries of $\rho$ using of the Maximum
Entropy principle. The last step implies the solution of a
nonlinear equation in order to shape the statistic to reproduce
exactly the moments obtained in the first estimation. Though this
last process may be delicate, we showed with simulated experiments
that it yields sound results when applied to coherent and number
states. Finally we demonstrated that the method exhibits a
sufficient robustness against errors deriving from the
Maximum-Likelihood estimation.
\section*{Acknowledgments}
MGAP is research fellow at \emph{ Collegio Alessandro Volta}. ARR
wishes to thank S. Olivares for fruitful discussions and helpful
suggestions.

\end{document}